\documentclass[twocolumn,prl]{revtex4}
\usepackage{graphicx}
\usepackage{bm}
\usepackage{amsmath}
\begin{document} \title{Spin-dependent properties of a two-dimensional
electron gas with ferromagnetic gates}

\author{C. Ciuti, J. P. McGuire, L. J. Sham}
\affiliation{Department of Physics, University of California San Diego,
La Jolla CA 92093-0319.}

\begin{abstract}  A theoretical prediction of the spin-dependent electron
self-energy and in-plane transport of a two-dimensional electron gas in
proximity with a ferromagnetic gate is presented. The application of the
predicted spin-dependent properties is illustrated by the proposal of a
device configuration with two neighboring ferromagnetic gates which
produces a magnetoresistance effect on the channel current generated by
nonmagnetic source and drain contacts.  Specific results are shown for a
silicon inversion layer with iron gates. The gate leakage current is
found to be beneficial to the spin effects.

\end{abstract}

\date{\today} \maketitle
%\draft

The generation, control, and measurement of electron spin polarization
are the essential ingredients of spintronics. The celebrated giant
magnetoresistance effect in metallic multilayers has lead to the
realization of remarkable non-volatile magnetic random access memories. A
great deal of interest is presently focused on {\it semiconductor}
spintronics because of the potentially appealing nonlinear
functionalities offered by semiconductors \cite{review}. One particularly
relevant system is that with planar confined carriers whose density can
be controlled by a gate.  Datta and Das \cite{Datta} proposed a
spintronics device consisting of a gated two-dimensional electron gas
(2DEG) with magnetic source and drain contacts.  The spin valve behavior
of the Datta and Das proposal depends on the orientation of the
magnetization in the source and drain contacts and is modulated by the
Rashba spin-orbit effect \cite{rashba} induced by the gate. Much recent
progress in spin injection \cite{review} and control of spin-orbit
interaction \cite{nitta} has been made towards the goal of an eventual
demonstration of the Datta and Das device.

In this letter, we investigate the spin-dependence of the energies and
lifetimes of a two-dimensional electron gas where the ferromagnetic
elements are the {\it gates}, while the source and drain are nonmagnetic
contacts. The spin-dependence of the electronic transport arises out of
the proximity to the ferromagnetic gates, in stark contrast to the
injection of spin-polarized electrons from the ferromagnetic source.
Recent experiments
\cite{FMimprinting,FMcoherence} have shown that large spontaneous spin
polarization can be achieved in an optically excited semiconductor layer
in contact with a ferromagnet . This phenomenon is explained in terms of
spin-dependent reflection at the semiconductor/ferromagnet interface
\cite{imprinting}.  For quantum confined semiconductor electrons, the
proximity effect with the ferromagnet is equivalent to multiple
reflections. A required condition is that the interface barrier is
sufficiently thin to allow for sizeable wave-function coupling. This is
analogous to the proximity effect of a metal such as Pd next to Fe
\cite{ivan}.  Our scheme avoids the restrictions of the Rashba effect,
namely narrow gap semiconductors (for large spin-orbit coupling) and
one-dimensional channels (to maximize the favorable directions of spin
torque). There is a wider range of appropriate systems for our scheme,
such as the inversion layer in Metal-Oxide-Semiconductor (MOS)
structures, the 2DEG in a modulation doped III-V quantum well, and the
InAs accumulation or inversion layers. The paper is structured as
follows. We shall discuss (i) the spin-dependent properties of a 2DEG
with a single ferromagnetic gate, (ii) a device proposal of a spin valve
with two neighboring gates, (iii) model results for a silicon inversion
layer with iron gates.

\begin{figure}[t!]
\includegraphics[width=8.3cm]{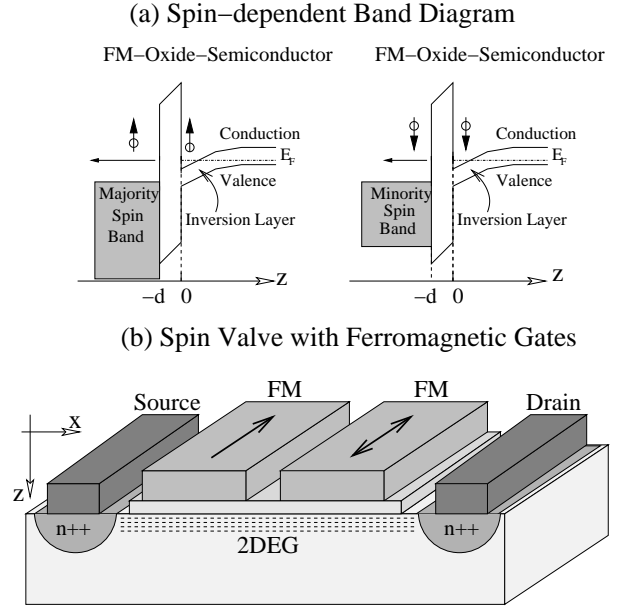}
\caption{(a) The spin-dependent band-diagram for a
ferromagnet-oxide-semiconductor junction (the bulk semiconductor is
p-doped). Left side: majority spin channel. Right side: minority spin.
(b) A schematic design of the proposed spin valve with two ferromagnetic
gates.  Source and drain are nonmagnetic contacts. }
\label{diagram}
\end{figure}

{\bf (i) 2DEG with a single ferromagnetic gate} \\ The tunneling coupling
of the electron wave function in the 2DEG to the exchange-split band
states in the metal produces a spin-dependent broadening and shift of the
quantized electron energy in the 2DEG. Namely, we have a complex energy
$E^{\rm s}_{\pm} = {\rm Re} (E^{\rm s}_{\pm}) - i \hbar / (2\tau_{\pm})$
where $E^{\rm s}_{+}$ ($E^{\rm s}_{-}$) refers to the majority (minority)
spin channel.  Note that
$\tau_{\pm}$ describes the escape time of the quasi-bound semiconductor
electrons into the ferromagnet band continuum.  Due to the spin-dependent
self-energy, the in-plane transport of the 2DEG is described by the
spin-dependent conductance $\sigma_{\pm}$ whose expression is
\begin{equation}
\frac{1}{\sigma_{\pm}} = \frac{m_{x}}{n_{\pm} e^2}~\left
(\frac{1}{\tau_0} +
\frac{1}{\tau_{\pm}} \right )~,
\end{equation} where $\tau_0$ is the ordinary spin-independent Drude
scattering time,
$m_x$ the in-plane effective mass and $n_{\pm}$ the spin-dependent
density of the 2DEG due to the spin splitting ${\rm Re}(E^{\rm
s}_{+}-E^{\rm s}_{-})$.
  However, this is not the only aspect of the problem, because a fraction
of the channel current is leaked into the gate (implying that drain and
source currents are different). The spin-dependent
leakage probabilities $\alpha_{\pm}$ for {\it non-equilibrium}
electrons injected into the channel are governed
by the {\it transit times} $ \tau_{\rm tr}^\pm = L_{\rm
G}/v_{x}^{\pm}$ where $L_{\rm G}$ is the length of the gate (along the
$x$-direction) and $v^{\pm}_{x}$ are the spin-dependent drift
velocities induced by the source-drain bias 
. Namely,
\begin{equation}
\alpha_{\pm} = \int^{\tau_{\rm tr}^\pm}_{0}dt~ \frac{1}{\tau_{\pm}}
~e^{-t/{\tau_{\pm}}} = 1 - \exp \left ( -{\frac{L_{\rm G}}{v_{x}^{\pm}
\tau_{\pm}}} \right )~.\label{probability}
\end{equation} In the linear regime, the drift velocity is $v_{x}^{\pm} =
\mu_{\pm} ~ {\mathcal E}_x$ where
${\mathcal E}_x$ is the in-plane field due to the source-drain bias
and $\mu_{\pm} = \sigma_{\pm}/(n_{\pm} e)$ are the spin-dependent mobilities.
When the lateral extension of the 2DEG channel is shorter than the
spin mean-free path, the majority and minority spin channels can be
considered decoupled in the active region between the source and drain
contacts.  Hence, the spin-dependent drain current will be
$j^{\rm D}_{\pm} \propto \sigma_{\pm} (1 - \alpha_{\pm}) =
\tilde{\sigma}_{\pm}$, i.e., the channel conductance $\sigma_{\pm}$ is
renormalized by the leakage probability. Note that the shorter the
$\tau_{\pm}$, the smaller are both $\sigma_{\pm}$ and $(1-
\alpha_{\pm})$. Hence, the leakage current, which ordinarily
reduces the channel current, actually enhances the difference between the
renormalized resistances
$1/\tilde{\sigma}_{\pm}$ of the two spin channels.

{\bf (ii) Spin valve with two ferromagnetic gates} \\ The
spin-dependent in-plane transport due to the proximity of a
ferromagnetic gate can be conveniently exploited to design a spin
valve (see Fig.~1b).  If two neighboring ferromagnetic gates are used,
the electron current detected at the drain contact will depend on the
relative orientation of the magnetizations of the two gates, giving
rise to the magnetoresistance ratio ${\mathcal R} = (j^{\rm D}_{++} -
j^{\rm D}_{+-})/j^{\rm D}_{++}$ where $j^{\rm D}_{++}$ and $j^{\rm
D}_{+-}$ are the total drain currents for {\it parallel} and {\it
antiparallel} magnetizations respectively.  For simplificity, we
consider the case where the two ferromagnetic gates are identical and
the effect of the channel region between the two gates on the current
is negligible.  The spin-dependent channel current density,
$j_{\pm}(x) = \sigma_{\pm}(x) \partial_x \phi_{\pm}(x)$ is driven by
the electrochemical potential $\phi_{\pm}(x)$ and is constrained by
the continuity equation $\partial_x j_{\pm} = - j_{\pm}/[v_x^{\pm}(x)
\tau_{\pm}(x)]$ accounting for the leakage of current into the
gates. Combining the two equations leads to a second order
differential equation for $\phi_{\pm}(x)$.  The boundary conditions
are (a) $\phi_+ = \phi_- = \phi_{source}$ ($\phi_{drain}$) at the edge
of the gate near the nonmagnetic source (drain) contact, (b)
$\phi_{\pm}(x)$ and $j_{\pm}(x)$ vary continuously along $x$. 
Under the first gate, $\sigma_{\pm}(x) = \sigma_{\pm}$. Under the second
gate, $\sigma_{\pm}(x) = \sigma_{\pm}$ in {\it parallel} configuration, but
$\sigma_{\pm}(x) = \sigma_{\mp}$ in the {\it antiparallel} case.  We
have solved exactly this equation both for the parallel and antiparallel
configuration. The analytical expression for the magnetoresistance ratio
${\mathcal R}$ is quite involved and will be presented in a longer
publication.  However, we have verified that a good approximation
is given by
\begin{equation} {\mathcal R} \simeq
\left[ \frac{   \sigma_+~(1-\alpha_{+}) - \sigma_- (1-\alpha_-) }
{\sigma_+~(1-\alpha_{+}) + \sigma_- (1-\alpha_-) }  \right]^2~.
\label{GMR}
\end{equation} This formula is obtained by a circuit diagram of the two
spin channels in parallel and by properly combining in series the
leakage-renormalized conductances $\tilde{\sigma}_{\pm}= \sigma_{\pm}
(1-\alpha_{\pm})$ for the two gates.

{\bf (iii) Model results for a silicon inversion layer} \\ Here we
present reasonable estimates for a ${\rm Si}/{\rm SiO}_2/{\rm Fe}$
heterostructure. We consider a free-electron Hamiltonian with two
exchange split spin bands in the ferromagnet \cite{slon}, an insulator
barrier and the electrostatic confinement potential for the inversion
layer. We will focus on the low temperature regime in which only the
first quantized subband in the inversion layer is filled and the
confining potential can be approximated with a triangular shape
(Fig.~\ref{diagram}a).  As suggested by the recent literature on
tunneling \cite{Si2,Si3}, the conservation of the in-plane wave-vector is
relaxed and, hence, a one-dimensional calculation can be carried out.
Within this simplified model, the spin-dependent wave-function
$\Psi_{\pm}(z)$ of the quasi-bound electron in the inversion layer can be
written in terms of special Airy functions ${\rm Ai}$ and ${\rm Bi}$.
For $z > 0$ (the semiconductor region),
\begin{equation}
\Psi_{\pm}(z) =  A_{\pm}{\rm Ai} \left [ k_{s} (z- \bar{z}_{\sigma})
\right ],
\end{equation} where $k^3_{\rm s} = 2 m_{\rm s} {e \mathcal E}_{\rm
s}/\hbar^2$ depends on the electric field ${\mathcal E}_{\rm s}$   in the
inversion layer and the semiconductor effective mass  $m_{\rm s}$. The
position
$\bar{z}_{\pm}$ represents the turning point. For $-d < z < 0$ (the
insulator barrier region),
\begin{equation}
\Psi_{\pm}(z) =
\left \{ B_{\pm} {\rm Ai} \left [ k_{\rm b}  (z - \hat{z}_{\pm}) \right ]
+ C_{\pm}  {\rm Bi} \left [ k_{\rm b} (z - \hat{z}_{\pm}) \right ]
\right \} ,
\end{equation} where $k^3_{\rm b} = 2 m_{\rm b} {e \mathcal E}_{\rm
b}/\hbar^2$ and
${\mathcal E}_{\rm b} = (\epsilon_{\rm s}/\epsilon_{\rm b}) {\mathcal
E}_{\rm s}$ is the electric field in the barrier region ($\epsilon_{\rm
b}$ and
$\epsilon_{\rm s}$ being the dielectric constants)  and
$\hat{z}_{\pm} = - (U_{\rm b}- E^{\rm s}_{\pm})/(e {\mathcal
E}_{\rm b})$ is the extrapolated turning point. The quantity
$E^{\rm s}_{\pm}$ is the quantization energy of the inversion layer
subband and has to be calculated self-consistently. Finally, for $z < -d$,
\begin{equation}
\Psi_{\pm}(z) = D_{\pm} ~\exp \left [ -i
\left ( k^{\rm fm}_{\pm} + i \frac{1} {l^{\rm fm}_{\pm}} \right )~z
\right ] ,
\end{equation} where $k^{\rm fm}_{\pm} = \sqrt{(k^{\rm F}_{\pm})^2+  2
m_{\rm fm}(e{\mathcal E}_{\rm b} d+E^{\rm s}_{\pm})/\hbar^2}$ in terms of
$k^{\rm F}_{\pm}$, the spin-dependent Fermi wave-vector and $m_{\rm fm}$,
the ferromagnet band mass. The term $l^{\rm fm}_{\pm}$ is the mean-free
path in the ferromagnet. The coefficients $A_{\pm}$,
$B_{\pm}$, $C_{\pm}$, $D_{\pm}$ are determined by the usual boundary
conditions at the interfaces $z=0$ and
$z=-d$.  The resultant secular equation for
$k \bar{z}_{\pm}$ has a complex solution, yielding the spin-dependent
{\it complex } energy $E^{\rm s}_{\pm} = k_{\rm s}
\bar{z}_{\pm}(\hbar^2 k_{\rm s}^2/2 m_{\rm z})$ in the inversion layer.
\begin{figure}[t!]
\includegraphics[width=8.3cm]{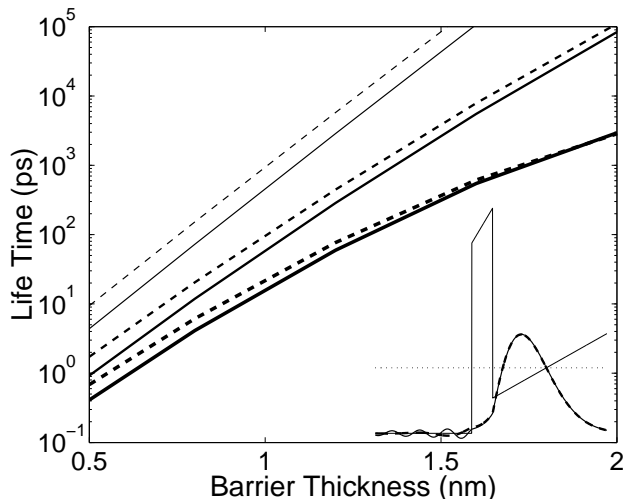}
\caption{Lifetimes (ps) of majority ($\tau_+$, solid lines) and
minority spin ($\tau_-$, dashed lines) as functions of the barrier
thickness (nm) for three different applied electric fields in the
oxide. Thin lines: ${\mathcal E}_{\rm b} = 2 \times 10^6$ V/cm. Medium
lines: ${\mathcal E}_{\rm b} = 7 \times 10^6$ V/cm. Thick lines:
${\mathcal E}_{\rm b} = 12 \times 10^6$ V/cm.  Parameters for ${\rm Si}$
($m_{\rm s} = 0.91 m_0$, $\epsilon_{\rm s} = 11.6$), ${\rm SiO_2}$
($m_{\rm b} = 0.3 m_0$, $\epsilon_{\rm b} = 3.8$, $U_{\rm b} = 3$ eV
\cite{Situnneling}), and ${\rm Fe}$ (Fermi wave-vectors $k^{\rm F}_{+}
= 1.1 ~\AA^{-1}$, $k^{\rm F}_{-} = 0.41 ~\AA^{-1}$ \cite{slon};
mean-free paths $l^{\rm fm}_{+} = 10 ~\AA$ and $l^{\rm fm}_{-} = 2.5~
\AA$\cite{Mills}).  The inset depicts the band diagram and the real
part of the spin-dependent wave-functions for the smallest barrier
thickness and the largest electric field.
\label{escapetimes} }
\end{figure}

  Fig.~\ref{escapetimes} depicts the majority and minority spin life
times $\tau^{\rm s}_+$ (solid lines) and $\tau^{\rm s}_-$ (dashed
lines) as functions of the oxide thickness.  The ultrathin range of
the oxide thickness and the oxide electron effective mass are based on
the recent measurements of electron tunneling across the silicon-based
MOS junction \cite{Situnneling,Si2,Si3}. The spin-dependent life time
decreases exponentially with decreasing barrier thickness. The ratio
of the times for two spins, $\tau^{\rm s}_-/\tau^{\rm s}_+$, is
dependent on the parameters of the ferromagnetic junction and can
exceed a factor of 2.  An increase of the electric field in the oxide
(via the gate voltage) dramatically shorten the lifetimes, approaching
the picosecond range.  The oxide electric fields ${\mathcal E}_{\rm
b}$ from the top to the bottom curves are 2, 7, and 12 $ \times 10^6$
V/cm (the highest value approaches the dielectric breakdown).  These
results show a significant spin-dependence of the channel mobility for
ultrathin oxide thickness.  The spin splitting (not shown) is
comparable to the energy broadening $(\hbar/2\tau^{\rm s}_+) +
(\hbar/2\tau^{\rm s}_-)$ and is about 0.3~meV when the lifetime is of
the order 1~ps.  Thus, for this specific system, the spin splitting
can be disregarded because it is much smaller than the Fermi energy of
the 2DEG.  The leakage probability $\alpha_{\pm}$ depends also on the
drift velocity according to Eq.~(\ref{probability}).  For a gate
length $L_{\rm G}= 0.1~\mu$m, in-plane field ${\mathcal E}_x = 0.2
~{\rm V}/\mu$m and Drude scattering time $\tau_0 = 1$ ps, the transit
time in the gated region is $\tau_{\rm tr} \approx 0.5 $ ps.  This means
that when the spin-dependent lifetimes are $\tau_{+} = 1$ ps and
$\tau_{-} = 2$ ps, the leakage probabilities are $\alpha_+ \approx 40
\%$ and $\alpha_- \approx 20 \%$. With these parameters, we get a
magnetoresistance ratio ${\mathcal R} \approx 10 \%$.  A decrease of
the drift velocity (through a decrease of the source-drain bias) will
make the leakage of the channel current more severe (thus reducing the
drain current), although it will increase ${\mathcal R}$ according to
Eq. (\ref{GMR}).

In summary, we have predicted the spin-dependent properties of a 2DEG
system due to the proximity of ferromagnetic gates, showing the potential
spintronics applications of this kind of configuration.

\begin{acknowledgments} This work is supported by DARPA/ONR
N0014-99-1-1096 and NSF DMR 0099572. CC is also grateful to the Swiss
National Science Foundation for additional support. JPM acknowledges
support by the California Institute for Telecommunications and Information
Technology. We thank Edward Yu, Daniel Schaadt, and Ivan Schuller for
helpful discussions.

\end{acknowledgments}

\end{document}